%% file: EAngelakis.tex
\documentclass{fmj2010}
\usepackage{natbib}
\usepackage{graphicx}

\def\fgamma{\textit{F}-GAMMA}
\def\myfgrst{\textit{Fermi}-GST}

\def\fermi{\textit{Fermi}}

\begin{document}
   \title{The \fgamma\ program: multi-wavelength AGN studies in the \myfgrst\ era}

   \author{E. Angelakis\inst{1}, L. Fuhrmann\inst{1},
     I. Nestoras\inst{1}, J. A. Zensus\inst{1}, N. Marchili\inst{1},
     V. Pavlidou\inst{2} 
          \and
          T. P. Krichbaum\inst{1}
          }

   \institute{Max-Planck-Institut f\"r Radioastronomie, Auf dem H\"gel 69, Bonn 53121,  Germany
         \and
             California Institute of Technology, 1200 East California
             Blvd., Pasadena CA 91125, USA    
             }

             \abstract{The \fgamma\ program is a coordinated effort of
               several observing facilities to understand the
               AGN/blazar phenomenon via a multi-frequency monitoring
               approach, especially in the era of \myfgrst. Some 60
               prominent sources are monitored monthly with the
               Effelsberg 100-m telescope, the IRAM 30-m telescope and
               more frequently but in a less uniform fashion at the
               APEX 12-m telescope, covering from 2.64 to
               345\,GHz. The program has been running since January
               2007 and here some of its findings are summarized. (a)
               There are two major variability patterns that the
               spectra of sources follow, one
               spectral-evolution-dominated and one achromatic. (b)
               The FSRQs show higher brightness temperatures
               indicative of larger Doppler factors at play and (c) a
               statistically significant radio-$\gamma$-ray
               correlation has been found with a method recently
               suggested by Pavlidou et al. (in prep.). }

             \authorrunning{Angelakis et al.}
             \titlerunning{The \fgamma\ program: multi-wavelength AGN studies in the \myfgrst\ era}

   \maketitle
%

\section{Introduction}
According to our current and rather complete understanding, in terms
of system configuration, of the magnificent phenomenon of Active
Galactic Nuclei (AGNs), the inconceivable amounts of energy radiated
by these systems are generated by a super-massive black hole living in
the nuclear region of the host galaxy. The even more extended variety
of AGN phenomenologies are attributed to the same system seen at
different viewing angles by the admittedly elegant unified model,
reviewed by \cite{Urry1995PASP}. Blazars, as the most dramatic
manifestation of the AGN phenomenon, display extreme observational
characteristics due to their close to our line-of-sight orientation
and the consequent boosting effects. Their double humped Spectral
Energy Distribution (SED) is explained by means of two components. The
low energy peaked one is explained as a synchrotron component whereas
the high energy peaked one is assumed to be the result of inverse
Compton processes either on external seed photons (from the disc, the
torus or the broad line region) or synchrotron photons. What the exact
mechanism is, is still unclear. Furthermore, what exactly causes the
variability at vast timescale ranges is also unknown and so are
numbers of fundamental questions.

Among the several incredible opportunities offered by \myfgrst\ is the
densely sampled $\gamma$-ray light curves and spectra it provides and
which for the first time allow really simultaneous multi-energy
studies. The \fgamma\ program is utilizing this approach for
understanding some of these fundamental questions. In particular,
roughly 60 sources have been monitored monthly since January 2007 with
the Effelsberg 100-m telescope, the IRAM 30-m telescope and the APEX
12-m telescope covering vast frequency range. The observations at
different telescopes are coordinated within 1 week. Here a few
findings are discussed after a short introduction to the program.

\section{The \fgamma\ program}
It has been discussed elsewhere
\citep{fuhrmann2007AIPC,angelakis2009arXiv0910} that the \fgamma\
program is the coordinated effort for a multi-frequency study of AGN
astrophysics in the light of \myfgrst\ gathering data since January
2007. The pivotal facilities are the Effelsberg 100-m telescope
covering the range between 2.64 and 43.00\,GHz in 8 frequency steps,
the IRAM \mbox{30-m} telescope covering 86, 142, 220\,GHz and the Apex
\mbox{12-m} telescope operating at 345\,GHz. The monitoring is done
monthly on a sample of ~60 sources. The observations at these
facilities are coordinated within roughly a week.  Along with this
effort there take place several other closely collaborating programs
such as the OVRO monitoring program at 15\,GHz, the IRAM \mbox{30-m} polarization
monitoring program (Thum et al. and \cite{agudo2010ApJS_189_1A}), and the
Perugia AIT blazar monitoring
\citep{tosti2002MmSAI,ciprini2008bves}. Here we present Effelsberg
data alone. The light curves produced and animated spectra are
constantly updated at www.mpifr.de/div/vlbi/fgamma.

\section{The sample}
The main motivation for this work has been the multi-band approach to
AGN physics and especially the blazar phenomenon, in particular the
investigation of correlations between the radio and $\gamma$-ray
emission. For this reason the first sample consisted of 61 blazars (32
FSRQs, 23 BL\,Lacs, 3 radio galaxies and 3 unclassified blazars)
selected mostly on their likelihood of being \myfgrst\ detectable
(earlier EGRET detections). The release of the LAT Bright AGN Source
list \citep[LBAS,][]{Abdo_2009ApJ_700_597A} showed that 29 of the 61
sources ($\sim 47\%$ ) were detected by LAT in the first three months
of operation.  In a major revision of the source sample, LBAS as well
as the \fermi-LAT First Source Catalog, were consulted
\citep{abdo_2010ApJS_188_405A} in order to build up a new source
sample with maximum \fermi-LAT detectability and reasonably fast and
intense variability. Currently the observed sample includes a total of
65 sources from which 60 are observed monthly. From these, 36 are
clarified as FSRQs, 17 as BL\,Lacs and 9 as unclassified blazars
\citep[classification
by][]{massaro2005,massaro2008,Massaro2009yCat_34950691M}. Additionally,
one source is classified as a Seyfert and three are radio galaxies.

From the above discussion it is very clear that the \fgamma\ sample
suffers severely from biases and therefore it is statistically
incomplete. Nevertheless, extracted generalizations may be tested by
careful comparison with other statistically complete samples. In any
case, any generalization must be stated with caution.


\section{Spectral Variability}
Blazar variability has been attributed to several factors and
accordingly certain models have been developed to interpret and
quantitatively describe it. One could distinguish between two model
categories: (a) models which attribute the variability to mechanisms
that impose spectral evolution on the observed events such as
shock-in-jet models \citep[][]{Marscher1985ApJ} or colliding
relativistic plasma shells \citep[][]{Guetta2004AnA} and (b) models
which explain the variability in terms of geometrical effects such as
systematic changes in the beam orientation \citep[e.g. lighthouse
effect, ][]{Camenzind1992AnA} which could for instance cause a change
in the Doppler factor and further induce achromatic changes in the
observed radio spectrum. The expected bi-modality in the
phenomenological behavior of the observed radio spectra is indeed seen
in the \fgamma\ data.

\begin{figure}[h!] 
\centering
\begin{tabular}{c}
\includegraphics[width=0.25\textwidth,angle=-90]{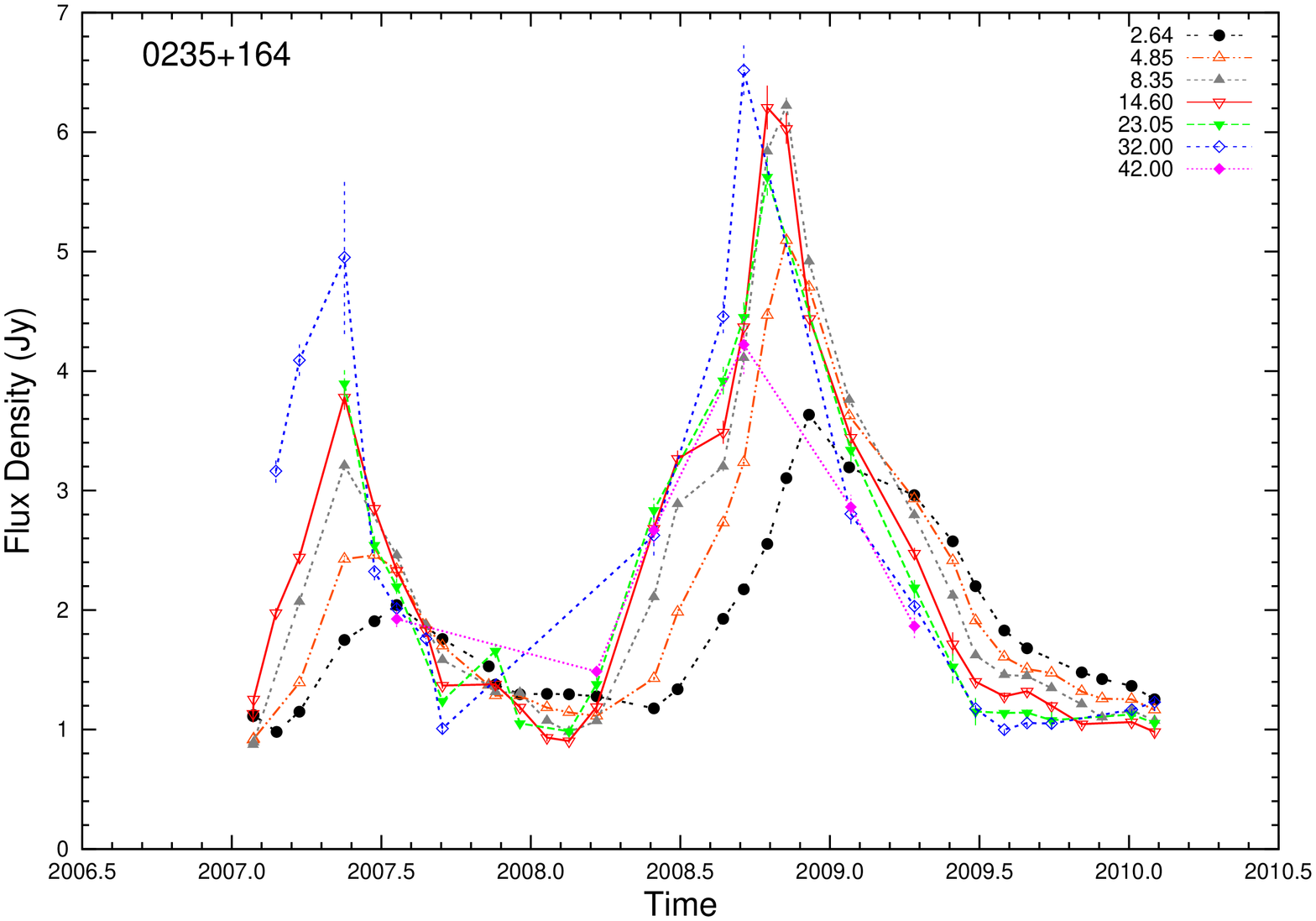} \\
\includegraphics[width=0.25\textwidth,angle=-90]{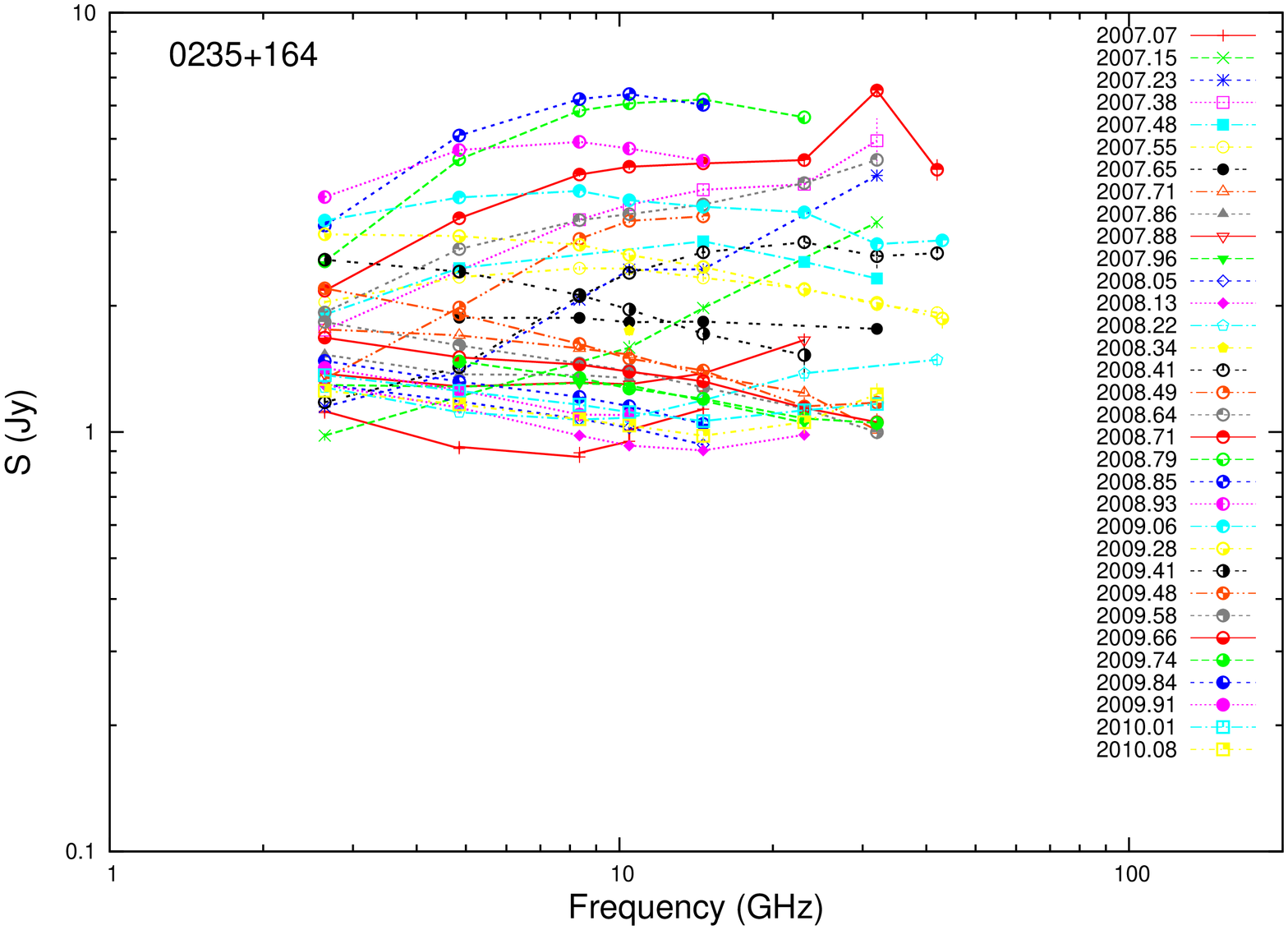}\\
\includegraphics[width=0.25\textwidth,angle=-90]{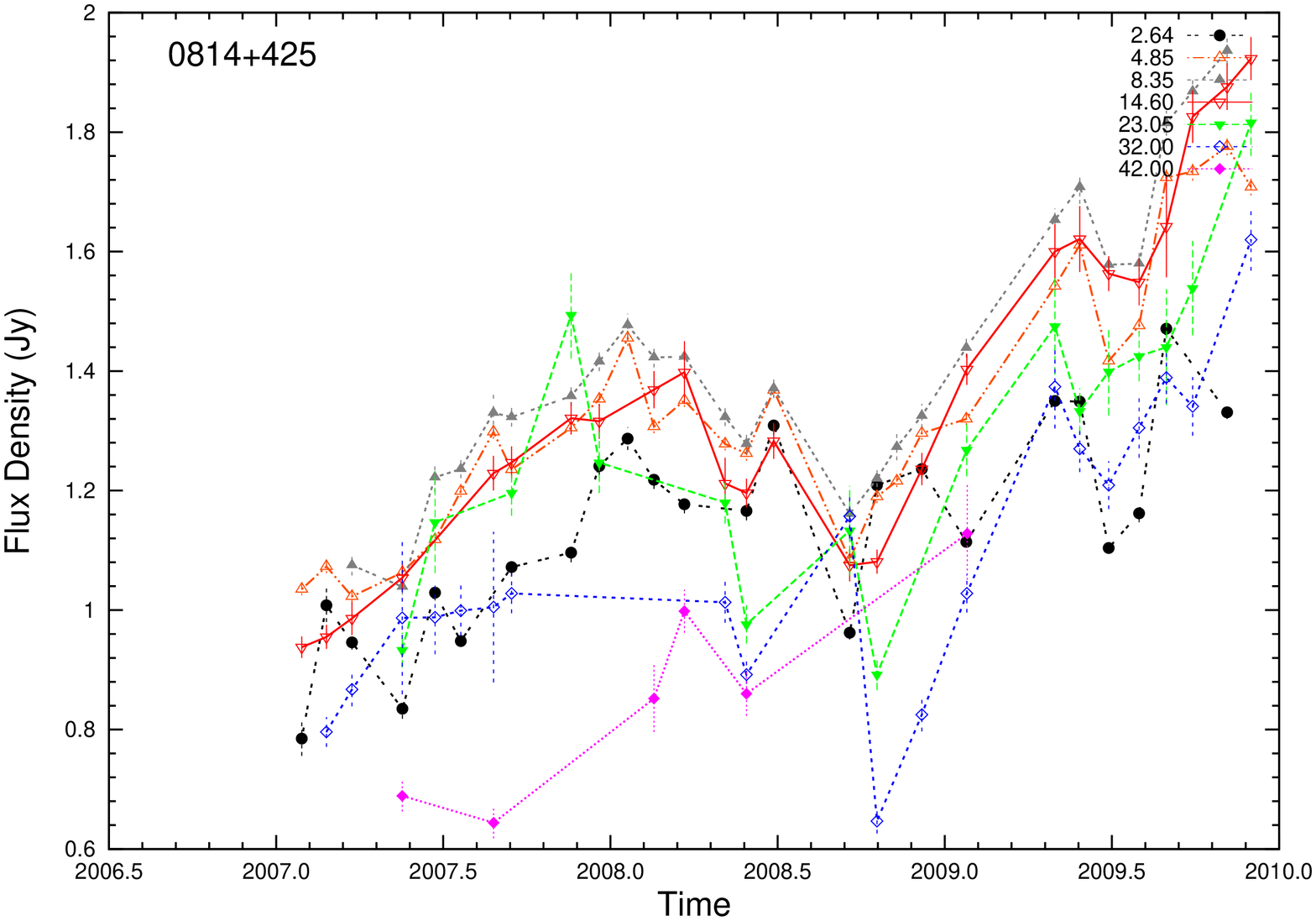} \\
\includegraphics[width=0.25\textwidth,angle=-90]{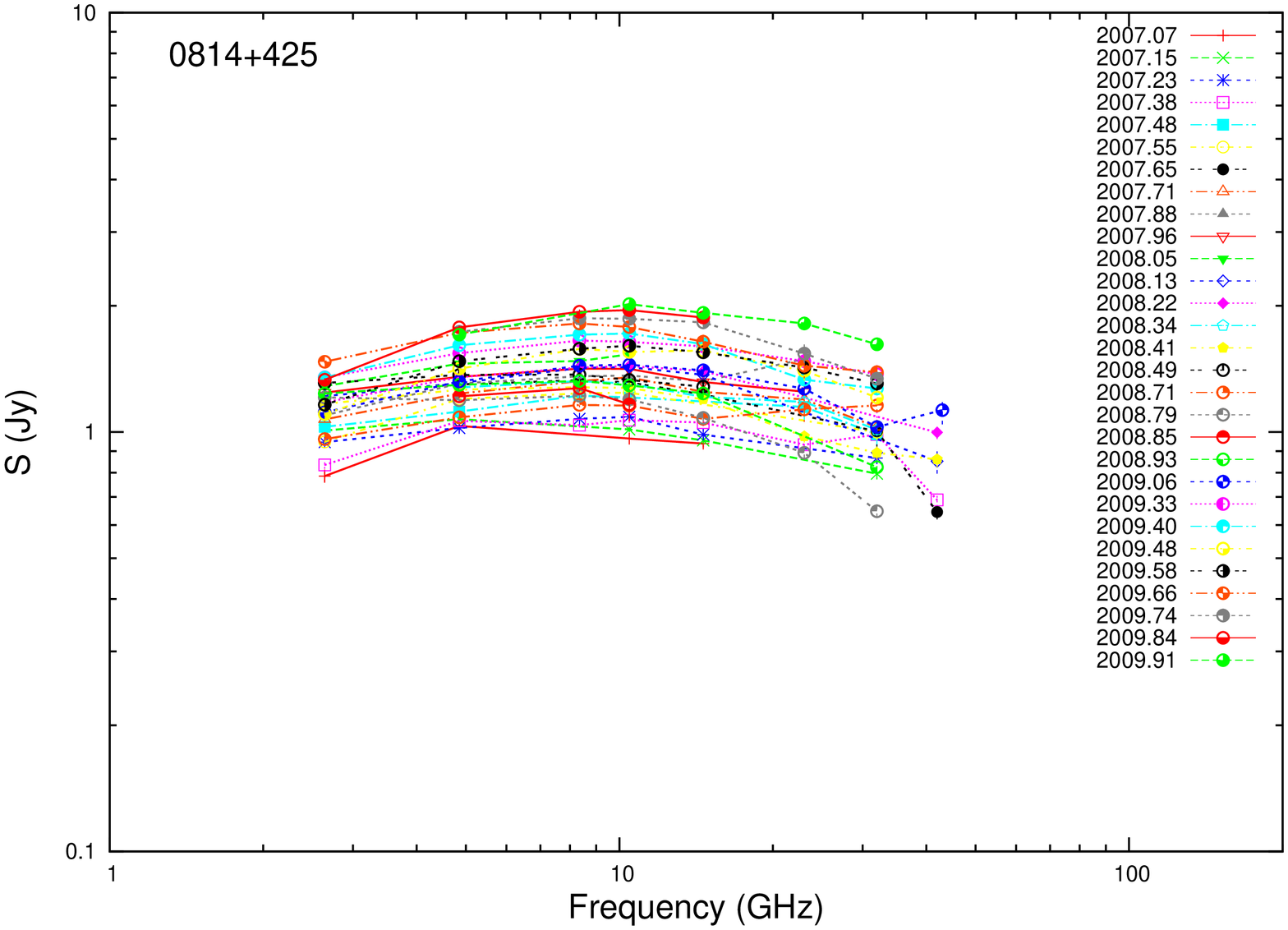}\\
\end{tabular}
\caption{Two prototype cases representing the two characteristic
  behaviors. In the case of $0235+164$ ($1^\mathrm{st}$ and
  $2^\mathrm{nd}$ panels from the top) the variability is dominated
  spectral evolution. The case of $0814+425$ is representative of
  achromatic variability indicating some geometrical effect.}
\label{fig1}
\end{figure}

Figure~\ref{fig1} shows the Effelsberg light curves and radio spectra
for two representative cases that resemble what would be expected from
the previous classification. In the case of 0235+164 the variability
is dominated by spectral evolution (hereafter {\sl type 1}), evidence of
a three-stage evolutionary path \citep[Compton, synchrotron and
adiabatic losses,][]{Marscher1985ApJ}. The case of $0814+425$ on the
other hand is representative of achromatic variability indicating some
geometrical effect (hereafter {\sl type 2}).

Interestingly, from the point of view of the variability pattern
followed by the observed spectra, all the sources fall in only five
classes which comprise modifications of these two basic behaviors.

\begin{figure}[h] 
\centering
\begin{tabular}{c}
\includegraphics[width=0.3\textwidth,angle=-90]{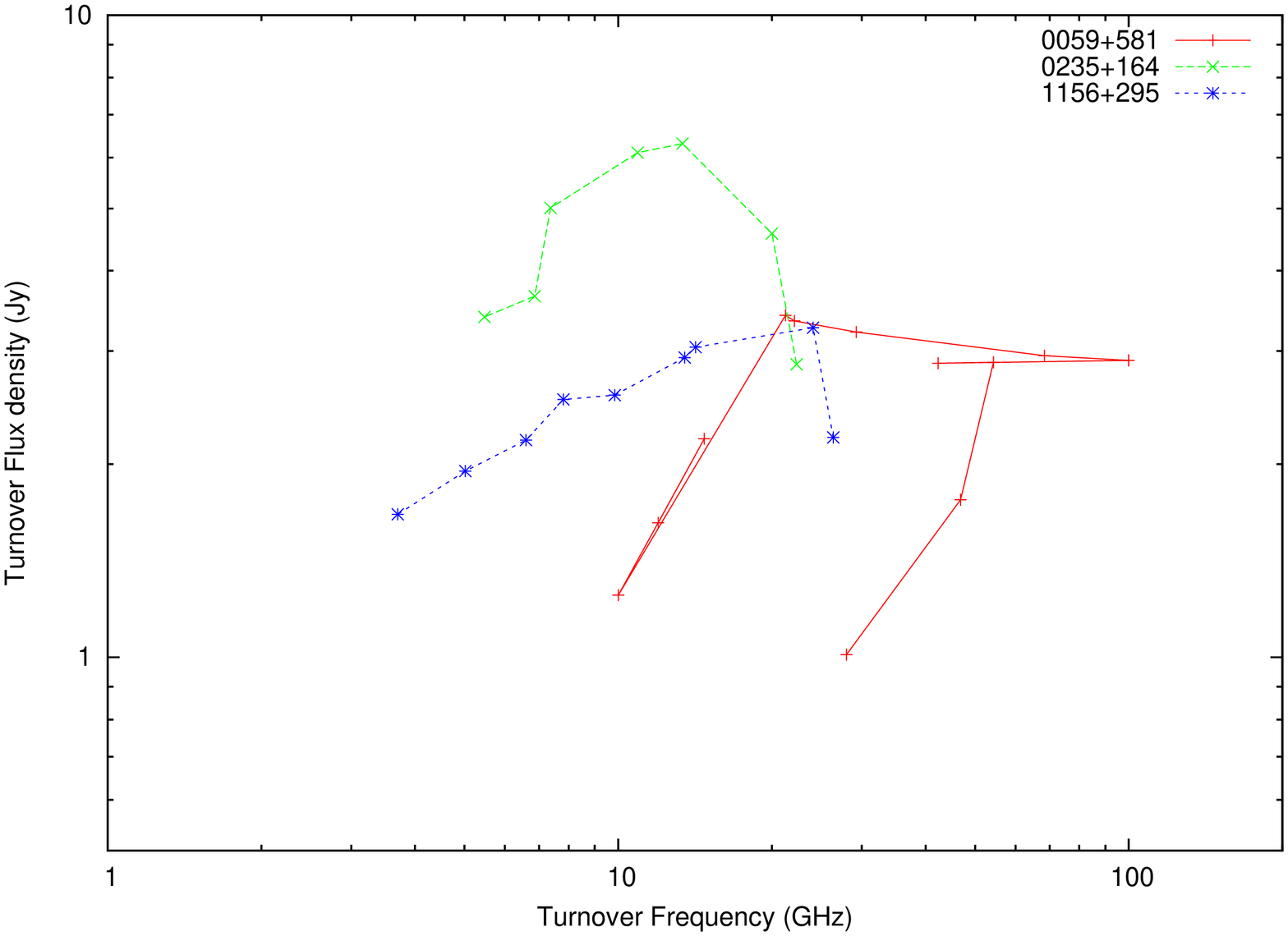} \\
\includegraphics[width=0.3\textwidth,angle=-90]{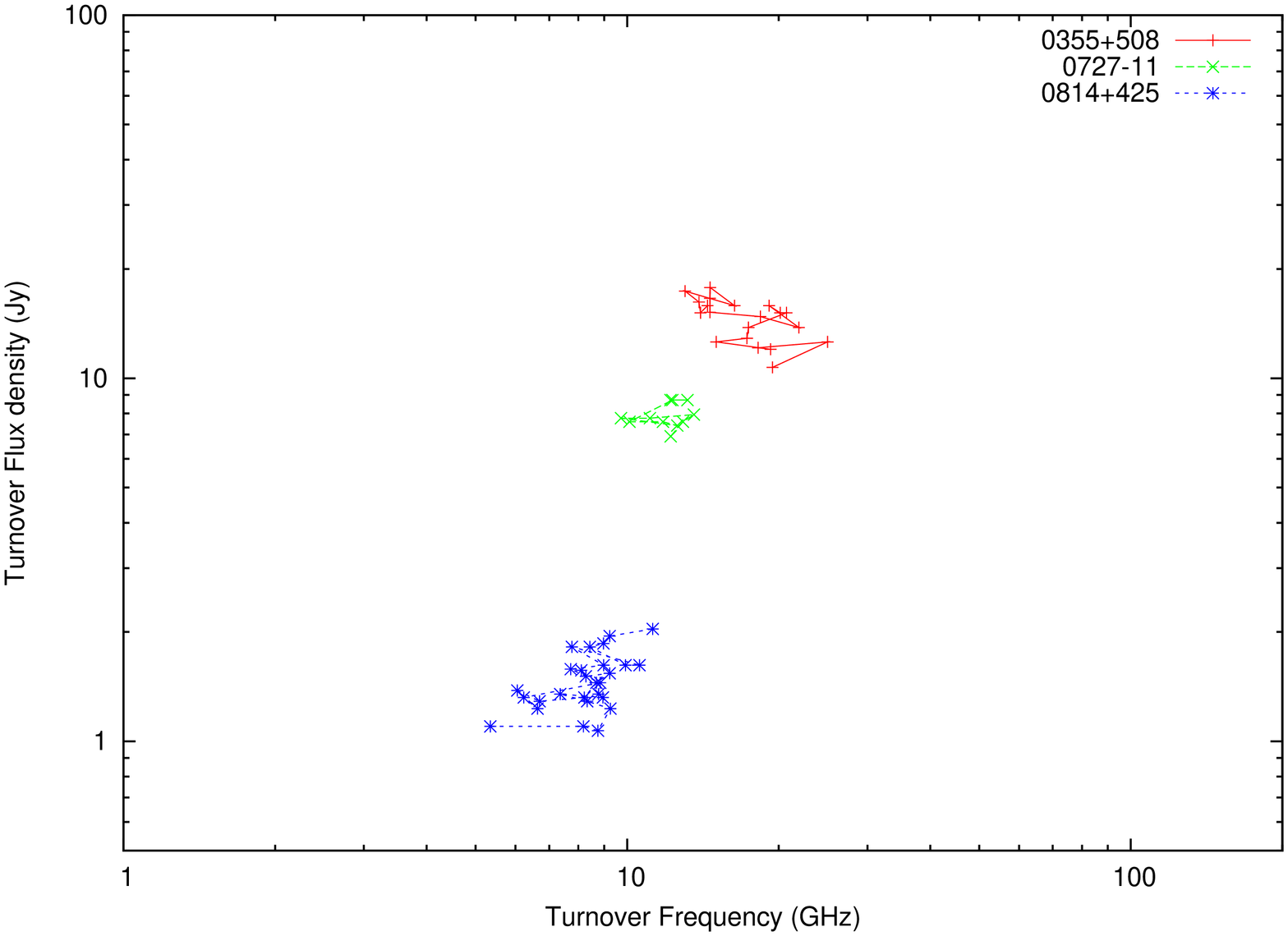}\\
\end{tabular}
\caption{The evolution of the peaks of convex spectra in the
  $S_{m}-\nu_{m}$ space for three members of each of the two
  characteristic classes {\sl type 1} and {\sl type 2}. In every case
  a quiescent spectrum of $-0.5$ is assumed ($S\propto
  \nu^{\alpha}$). Members of the same class follow qualitatively the
  same evolutionary paths.}
\label{fig2}
\end{figure}

The fact that there must be a fundamental difference in the mechanism
causing the variability becomes evident also from the evolutionary
paths followed by the turnover frequency and flux density at that
point ($S_\mathrm{m}$, $\nu_\mathrm{m}$). In figure~\ref{fig2} are
shown the evolutionary tracks for three cases of {\sl type 1} and
three cases of {\sl type 2} after the subtraction of mean quiescent
spectra (with $S\propto \nu^\alpha$ and $\alpha=-0.5$). The former
case seems to be described well by \cite{Marscher1985ApJ} whereas {\sl
  type 2}, needs a different interpretation. 
In a forthcoming publication (Angelakis et al. in prep.) we use this
approach to estimate source parameters, and investigate the
presence of possible quasi-periodicities in {\sl type 2} sources

\begin{figure}[h] 
\centering
\includegraphics[angle=-90, clip=true, width=0.4\textwidth]{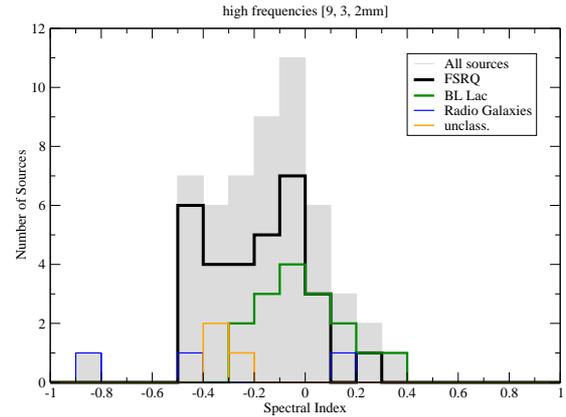} \\
\caption{The distribution of high-frequency spectral indices for 
BL\,Lacs and FSRQs with the former centering around flatter values than
  the latter.}
\label{fig4}
\end{figure}

From average spectra a low and high frequency spectral index can be
calculated in order to examine whether there is a significant
differentiation between BL\,Lacs and FSRQs. Figure~\ref{fig4} shows
the spectral indices between 9\,mm observed with the Effelsberg
telescope and 3 and 2\,mm obtained with the IRAM \mbox{30-m}
telescope. What seems to be the case is a tendency of the BL\,Lacs to
center around flatter high-frequency spectral indices as compared to
FSRQs (and is indicated by statistical tests). This may for instance
be due to the fact that the BL\,Lacs systematically peak at higher
frequencies or, alternatively, their flares expand before reaching
lower frequencies, maybe due to lower power. This requires further
investigations.


\section{Time series analysis}
The level of variability present at all frequencies and in all sources
becomes obvious as {\sl excess variance} and has been quantified via a
formal $\chi^2$ test for which a significance reference level of
99.9\,\% has been adopted. For the first 2.5 years of observations
(that is for the first target sample), $\chi^2$ tests show that almost
all the sources appear significantly variable at all wavelengths
(91\,\% of the sources are significantly variable with the proportion
of variable sources dropping towards higher frequencies due to larger
measurement uncertainties at these frequencies). Using the mean {\sl
  rms} ($\left< \sigma\right >$) averaged over all sources at each
frequency as a measure of the variability amplitude, one can clearly
see a monotonically rising trend as a function of frequency. That is
expected from sources which are mostly spectral-evolution-dominated.

\begin{figure}[h] 
\centering
\includegraphics[width=0.4\textwidth,angle=-90]{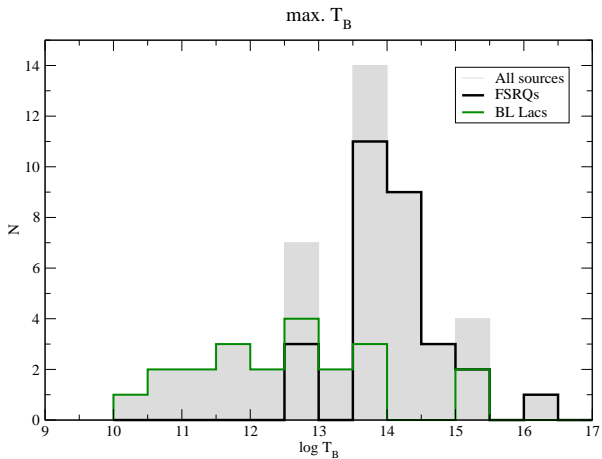} 
\caption{The brightness temperature distribution for FSRQs and BL\,Lacs
separately.}
\label{fig3}
\end{figure}

In order to investigate the presence of characteristic timescales in
the acquired light curves, the first order {\sl Structure Function}
has been used
\citep{Rutman1978IEEEP_66_1048R,Paltani1997AnA_327_539P,Simonetti1985ApJ}. From
the time series analysis applied to the first 2.5 years of data, the
variability timescales range between 80 and 500 days. From the
estimated timescales and on the basis of some fundamental assumptions
(causality, a single emitting component), one can estimate the
brightness temperatures $T_\mathrm{b} = 4.5\cdot 10^{10} \Delta S
\left (\frac{\lambda D_\mathrm{L}}{\tau (1+z)^2}\right )^2$, with $T_\mathrm{b}$ in
K, $\Delta S$ the flux density variations in Jy, $\lambda$ in cm, $D_{\mathrm{L}}$
the luminosity distance in Mpc, $\tau$ the characteristic timescale in
days and $z$ the redshift. Figure~\ref{fig3} shows the maximum
brightness temperature distribution (calculated from the fastest time
scales reliably detected) for FSRQs and BL\,Lacs. There is a clear and
statistically significant separation between the two classes, with the
FSRQs showing brightness temperatures systematically higher than the
BL\,Lacs. A possible explanation for this could be that the former
undergo stronger Doppler boosting.

\section{Radio versus $\gamma$-ray fluxes}
One of the remaining fundamental questions in AGN astrophysics is
where and how high energy emission is produced. Searches for
correlations between radio and $\gamma$-ray luminosities is used to
study the connection between the mechanisms producing them and has
been a rather debatable field. The situation becomes even more
perplexing due to artifacts that may be introduced by selection
biases, redshift dependencies, lack of simultaneity and so forth.

\begin{figure}[h] 
\centering
\includegraphics[width=0.4\textwidth,angle=-90]{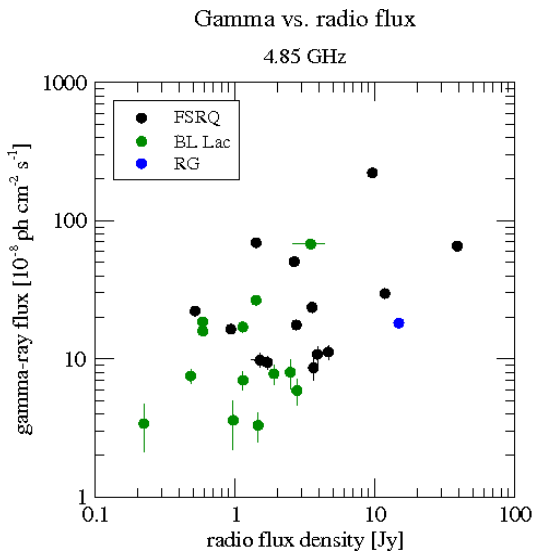} 
\caption{Flux-flux correlation plots with three-month averaged radio
  data at 4.85\,GHz.}
\label{fig5}
\end{figure}

For a sample of 29 \fgamma\ sources from the first sample detected by
\myfgrst, simultaneous $\gamma$-ray and radio data have been used to
investigate the presence of such a correlation. Figure~\ref{fig5}
shows that there indeed exists a correlation. A new Monte-Carlo method
has been used to assess its significance and it shows that it is
indeed intrinsic. The details of the newly applied method are describe
by Pavlidou et al. (in prep.).

\section{Conclusions}

After 3.5 years of observing, the \fgamma\ program has produced a large
volume of data which allows very detailed blazar studies. Some of them
have been discussed here:
\begin{itemize}
\item The spectrum variability is not erratic but follows only two
  behaviors: (a) one dominated by spectral evolution and (b) one
  achromatic variability pattern resembling rather a geometrically
  induced variability.  All the sources fall in these categories and
  their modifications. Differences in the variability properties of
  sources from different classes must be sought.
\item The high-frequency spectral indices imply a separation between
  FSRQs and BL\,Lacs, possibly due to the BL\,Lacs peaking at higher
  frequencies.
\item A clear distinction in the brightness temperature distributions
  for BL\,Lacs and FSRQs may be due to higher Doppler factors at play
  in FSRQs.
\item A newly suggested Monte-Carlo method assesses statistically
  significant radio and gamma-ray flux correlations. 
\end{itemize}

\begin{acknowledgements}
  Based on observations with the 100-m telescope of the MPIfR
  (Max-Planck-Institut f\"ur Radioastronomie) at Effelsberg. IN is a
  member of the International Max Planck Research School (IMPRS) for
  Astronomy and Astrophysics at the Universities of Bonn and
  Cologne. This work has made use of observations with the IRAM 30-m
  telescope.
\end{acknowledgements}


\end{document}

%% file: EAngelakis.bbl
\begin{thebibliography}{17}
\expandafter\ifx\csname natexlab\endcsname\relax\def\natexlab#1{#1}\fi

\bibitem[{{Abdo} {et~al.}(2010){Abdo}, {Ackermann}, {Ajello}, {Allafort},
  {Antolini}, {Atwood}, {Axelsson}, {Baldini}, {Ballet}, {Barbiellini},
  {Bastieri}, {Baughman}, {Bechtol}, {Bellazzini}, {Belli}, {Berenji},
  {Bisello}, {Blandford}, {Bloom}, {Bonamente}, {Bonnell}, {Borgland},
  {Bouvier}, {Bregeon}, {Brez}, {Brigida}, {Bruel}, {Burnett}, {Busetto},
  {Buson}, {Caliandro}, {Cameron}, {Campana}, {Canadas}, {Caraveo}, {Carrigan},
  {Casandjian}, {Cavazzuti}, {Ceccanti}, {Cecchi}, {{\c C}elik}, {Charles},
  {Chekhtman}, {Cheung}, {Chiang}, {Cillis}, {Ciprini}, {Claus},
  {Cohen-Tanugi}, {Conrad}, {Corbet}, {Davis}, {DeKlotz}, {den Hartog},
  {Dermer}, {de Angelis}, {de Luca}, {de Palma}, {Digel}, {Dormody}, {Silva},
  {Drell}, {Dubois}, {Dumora}, {Fabiani}, {Farnier}, {Favuzzi}, {Fegan},
  {Ferrara}, {Focke}, {Fortin}, {Frailis}, {Fukazawa}, {Funk}, {Fusco},
  {Gargano}, {Gasparrini}, {Gehrels}, {Germani}, {Giavitto}, {Giebels},
  {Giglietto}, {Giommi}, {Giordano}, {Giroletti}, {Glanzman}, {Godfrey},
  {Grenier}, {Grondin}, {Grove}, {Guillemot}, {Guiriec}, {Gustafsson},
  {Hadasch}, {Hanabata}, {Harding}, {Hayashida}, {Hays}, {Healey}, {Hill},
  {Horan}, {Hughes}, {Iafrate}, {J{\'o}hannesson}, {Johnson}, {Johnson},
  {Johnson}, {Johnson}, {Kamae}, {Katagiri}, {Kataoka}, {Kawai}, {Kerr},
  {Kn{\"o}dlseder}, {Kocevski}, {Kuss}, {Lande}, {Landriu}, {Latronico}, {Lee},
  {Lemoine-Goumard}, {Lionetto}, {Llena Garde}, {Longo}, {Loparco}, {Lott},
  {Lovellette}, {Lubrano}, {Madejski}, {Makeev}, {Marangelli}, {Marelli},
  {Massaro}, {Mazziotta}, {McConville}, {McEnery}, {Michelson}, {Minuti},
  {Mitthumsiri}, {Mizuno}, {Moiseev}, {Mongelli}, {Monte}, {Monzani},
  {Moretti}, {Morselli}, {Moskalenko}, {Murgia}, {Nakajima}, {Nakamori},
  {Naumann-Godo}, {Nolan}, {Norris}, {Nuss}, {Ohno}, {Ohsugi}, {Omodei},
  {Orlando}, {Ormes}, {Ozaki}, {Paccagnella}, {Paneque}, {Panetta}, {Parent},
  {Pelassa}, {Pepe}, {Pesce-Rollins}, {Pinchera}, {Piron}, {Porter}, {Poupard},
  {Rain{\`o}}, {Rando}, {Ray}, {Razzano}, {Razzaque}, {Rea}, {Reimer},
  {Reimer}, {Reposeur}, {Ripken}, {Ritz}, {Rochester}, {Rodriguez}, {Romani},
  {Roth}, {Sadrozinski}, {Salvetti}, {Sanchez}, {Sander}, {Saz Parkinson},
  {Scargle}, {Schalk}, {Scolieri}, {Sgr{\`o}}, {Shaw}, {Siskind}, {Smith},
  {Smith}, {Spandre}, {Spinelli}, {Starck}, {Stephens}, {Striani}, {Strickman},
  {Strong}, {Suson}, {Tajima}, {Takahashi}, {Takahashi}, {Tanaka}, {Thayer},
  {Thayer}, {Thompson}, {Tibaldo}, {Tibolla}, {Tinebra}, {Torres}, {Tosti},
  {Tramacere}, {Uchiyama}, {Usher}, {Van Etten}, {Vasileiou}, {Vilchez},
  {Vitale}, {Waite}, {Wallace}, {Wang}, {Watters}, {Winer}, {Wood}, {Yang},
  {Ylinen}, \& {Ziegler}}]{abdo_2010ApJS_188_405A}
{Abdo}, A.~A., {Ackermann}, M., {Ajello}, M., {et~al.} 2010, \apjs, 188, 405

\bibitem[{{Abdo} {et~al.}(2009){Abdo}, {Ackermann}, {Ajello}, {Atwood},
  {Axelsson}, {Baldini}, {Ballet}, {Barbiellini}, {Bastieri}, {Baughman},
  {Bechtol}, {Bellazzini}, {Blandford}, {Bloom}, {Bonamente}, {Borgland},
  {Bouvier}, {Bregeon}, {Brez}, {Brigida}, {Bruel}, {Burnett}, {Caliandro},
  {Cameron}, {Caraveo}, {Casandjian}, {Cavazzuti}, {Cecchi}, {Charles},
  {Chekhtman}, {Chen}, {Cheung}, {Chiang}, {Ciprini}, {Claus}, {Cohen-Tanugi},
  {Colafrancesco}, {Collmar}, {Cominsky}, {Conrad}, {Costamante}, {Cutini},
  {Dermer}, {de Angelis}, {de Palma}, {Digel}, {do Couto e Silva}, {Drell},
  {Dubois}, {Dumora}, {Farnier}, {Favuzzi}, {Fegan}, {Ferrara}, {Finke},
  {Focke}, {Foschini}, {Frailis}, {Fuhrmann}, {Fukazawa}, {Funk}, {Fusco},
  {Gargano}, {Gasparrini}, {Gehrels}, {Germani}, {Giebels}, {Giglietto},
  {Giommi}, {Giordano}, {Giroletti}, {Glanzman}, {Godfrey}, {Grenier},
  {Grondin}, {Grove}, {Guillemot}, {Guiriec}, {Hanabata}, {Harding}, {Hartman},
  {Hayashida}, {Hays}, {Healey}, {Horan}, {Hughes}, {J{\'o}hannesson},
  {Johnson}, {Johnson}, {Johnson}, {Johnson}, {Kadler}, {Kamae}, {Katagiri},
  {Kataoka}, {Kerr}, {Kn{\"o}dlseder}, {Kocian}, {Kuehn}, {Kuss}, {Lande},
  {Latronico}, {Lemoine-Goumard}, {Longo}, {Loparco}, {Lott}, {Lovellette},
  {Lubrano}, {Madejski}, {Makeev}, {Massaro}, {Mazziotta}, {McConville},
  {McEnery}, {McGlynn}, {Meurer}, {Michelson}, {Mitthumsiri}, {Mizuno},
  {Moiseev}, {Monte}, {Monzani}, {Moretti}, {Morselli}, {Moskalenko}, {Murgia},
  {Nolan}, {Norris}, {Nuss}, {Ohsugi}, {Omodei}, {Orlando}, {Ormes}, {Ozaki},
  {Paneque}, {Panetta}, {Parent}, {Pelassa}, {Pepe}, {Pesce-Rollins}, {Piron},
  {Porter}, {Rain{\`o}}, {Rando}, {Razzano}, {Razzaque}, {Reimer}, {Reimer},
  {Reposeur}, {Reyes}, {Ritz}, {Rochester}, {Rodriguez}, {Romani}, {Ryde},
  {Sadrozinski}, {Sanchez}, {Sander}, {Saz Parkinson}, {Scargle}, {Schalk},
  {Sellerholm}, {Sgr{\`o}}, {Shaw}, {Smith}, {Smith}, {Spandre}, {Spinelli},
  {Starck}, {Strickman}, {Suson}, {Tajima}, {Takahashi}, {Takahashi}, {Tanaka},
  {Taylor}, {Thayer}, {Thayer}, {Thompson}, {Tibaldo}, {Torres}, {Tosti},
  {Tramacere}, {Uchiyama}, {Usher}, {Vilchez}, {Villata}, {Vitale}, {Waite},
  {Winer}, {Wood}, {Ylinen}, \& {Ziegler}}]{Abdo_2009ApJ_700_597A}
{Abdo}, A.~A., {Ackermann}, M., {Ajello}, M., {et~al.} 2009, \apj, 700, 597

\bibitem[{{Agudo} {et~al.}(2010){Agudo}, {Thum}, {Wiesemeyer}, \&
  {Krichbaum}}]{agudo2010ApJS_189_1A}
{Agudo}, I., {Thum}, C., {Wiesemeyer}, H., \& {Krichbaum}, T.~P. 2010, \apjs,
  189, 1

\bibitem[{{Angelakis} {et~al.}(2009){Angelakis}, {Fuhrmann}, {Zensus},
  {Nestoras}, {Marchili}, {Krichbaum}, {Ungerechts}, {Max-Moerbeck},
  {Pavlidou}, {Pearson}, {Readhead}, {Richards}, \&
  {Stevenson}}]{angelakis2009arXiv0910}
{Angelakis}, E., {Fuhrmann}, L., {Zensus}, J.~A., {et~al.} 2009, ArXiv e-prints

\bibitem[{{Camenzind} \& {Krockenberger}(1992)}]{Camenzind1992AnA}
{Camenzind}, M. \& {Krockenberger}, M. 1992, \aap, 255, 59

\bibitem[{{Ciprini} {et~al.}(2008){Ciprini}, {Tosti}, {Nucciarelli},
  {Bagaglia}, {Fiorucci}, {Impiombato}, {Luciani}, {Maffei}, {Marchili},
  {Pascolini}, {Picarelli}, {Rizzi}, \& {Roncella}}]{ciprini2008bves}
{Ciprini}, S., {Tosti}, G., {Nucciarelli}, G., {et~al.} 2008, in Blazar
  Variability across the Electromagnetic Spectrum

\bibitem[{{Fuhrmann} {et~al.}(2007){Fuhrmann}, {Zensus}, {Krichbaum},
  {Angelakis}, \& {Readhead}}]{fuhrmann2007AIPC}
{Fuhrmann}, L., {Zensus}, J.~A., {Krichbaum}, T.~P., {Angelakis}, E., \&
  {Readhead}, A.~C.~S. 2007, in American Institute of Physics Conference
  Series, Vol. 921, The First GLAST Symposium, ed. S.~{Ritz}, P.~{Michelson},
  \& C.~A. {Meegan}, 249--251

\bibitem[{{Guetta} {et~al.}(2004){Guetta}, {Ghisellini}, {Lazzati}, \&
  {Celotti}}]{Guetta2004AnA}
{Guetta}, D., {Ghisellini}, G., {Lazzati}, D., \& {Celotti}, A. 2004, \aap,
  421, 877

\bibitem[{{Marscher} \& {Gear}(1985)}]{Marscher1985ApJ}
{Marscher}, A.~P. \& {Gear}, W.~K. 1985, \apj, 298, 114

\bibitem[{{Massaro} {et~al.}(2009){Massaro}, {Giommi}, {Leto}, {Marchegiani},
  {Maselli}, {Perri}, {Piranomonte}, \& {Sclavi}}]{Massaro2009yCat_34950691M}
{Massaro}, E., {Giommi}, P., {Leto}, C., {et~al.} 2009, VizieR Online Data
  Catalog, 349, 50691

\bibitem[{{Massaro} {et~al.}(2005){Massaro}, {Sclavi}, {Giommi}, {Perri}, \&
  {Piranomonte}}]{massaro2005}
{Massaro}, E., {Sclavi}, S., {Giommi}, P., {Perri}, M., \& {Piranomonte}, S.
  2005, Aracne, Roma, I

\bibitem[{{Massaro} {et~al.}(2008){Massaro}, {Sclavi}, {Giommi}, {Perri}, \&
  {Piranomonte}}]{massaro2008}
{Massaro}, E., {Sclavi}, S., {Giommi}, P., {Perri}, M., \& {Piranomonte}, S.
  2008, Aracne, Roma, I

\bibitem[{{Paltani} {et~al.}(1997){Paltani}, {Courvoisier}, {Blecha}, \&
  {Bratschi}}]{Paltani1997AnA_327_539P}
{Paltani}, S., {Courvoisier}, T., {Blecha}, A., \& {Bratschi}, P. 1997, \aap,
  327, 539

\bibitem[{{Rutman}(1978)}]{Rutman1978IEEEP_66_1048R}
{Rutman}, J. 1978, IEEE Proceedings, 66, 1048

\bibitem[{{Simonetti} {et~al.}(1985){Simonetti}, {Cordes}, \&
  {Heeschen}}]{Simonetti1985ApJ}
{Simonetti}, J.~H., {Cordes}, J.~M., \& {Heeschen}, D.~S. 1985, \apj, 296, 46

\bibitem[{{Tosti} {et~al.}(2002){Tosti}, {Ciprini}, \&
  {Nucciarelli}}]{tosti2002MmSAI}
{Tosti}, G., {Ciprini}, S., \& {Nucciarelli}, G. 2002, Memorie della Societa
  Astronomica Italiana, 73, 1024

\bibitem[{{Urry} \& {Padovani}(1995)}]{Urry1995PASP}
{Urry}, C.~M. \& {Padovani}, P. 1995, \pasp, 107, 803

\end{thebibliography}
